# Characterization of the Schottky Barrier in SrRuO$_3$/Nb:SrTiO$_3$ Junctions

Y. Hikita[a)], Y, Kozuka, T. Susaki, H. Takagi, and H. Y. Hwang[b)]

*Department of Advanced Materials Science, University of Tokyo, Kashiwa, Chiba 277-8561,*

*Japan*

Internal photoemission spectroscopy was used to determine the Schottky barrier height in rectifying SrRuO$_3$/Nb-doped SrTiO$_3$ junctions for 0.01 wt % and 0.5 wt % Nb concentrations. Good agreement was obtained with the barrier height deduced from capacitance-voltage measurements, provided that a model of the nonlinear permittivity of SrTiO$_3$ was incorporated in extrapolating the built-in potential, particularly for high Nb concentrations. Given the generic polarizability of perovskites under internal/external electric fields, internal photoemission provides a valuable independent probe of the interface electronic structure.

a) Electronic mail: kk47114@mail.ecc.u-tokyo.ac.jp

b) Also at: Japan Science and Technology Agency, Kawaguchi 332-0012, Japan.

Transition metal oxides with the perovskite structure have been widely investigated for their rich and exotic physical properties. With a perspective toward functional devices, much effort has been put into the study of heterojunctions of these materials. Recent examples include junctions which incorporate ferroelectricity,[1] resistive switching,[2,3] photocarrier injection,[4] and magnetic field sensitivity.[5] A common need in all cases is a quantitative understanding of the electronic structure at the interface. In particular, the most fundamental characteristic of rectifying metal-semiconductor junctions is the Schottky barrier height (SBH), which is the energy discontinuity between the Fermi level of the metal and the conduction band minimum (valence band maximum) of the n-type (p-type) semiconductor. Although the SBH can be estimated from current-voltage ($I$-$V$) and capacitance-voltage ($C$-$V$) measurements,[6] these measures have significant uncertainty arising from limited understanding of the dominant transport mechanisms,[7] as well as unusual features arising from incorporating materials with strongly correlated electrons and nonlinear permittivity.

Internal photoemission (IPE) is a direct, reliable method for measuring the SBH and band discontinuities in semiconductor heterostructures.[8,9] Using IPE, the SBH is determined by monitoring the photocurrent across the junction while illuminated by tunable monochromatic light. The SBH is detected as the threshold photon energy for which electrons can be photo-excited over the barrier. An advantage of this technique is that the SBH can be determined without external bias, probing the junction in its equilibrium state and eliminating the complexities related to an externally applied potential. This is particularly important for junctions with Nb-doped SrTiO$_3$ (Nb:SrTiO$_3$), a heavily utilized *n*-type semiconductor in transition metal oxide heterostructures, since the permittivity of SrTiO$_3$ varies dramatically at high electric fields.[10]

Here we examine the Schottky barrier formed between metallic SrRuO$_3$ and Nb:SrTiO$_3$ substrates. This interface is attractive for initial studies of IPE in oxide heterojunctions for several reasons. SrRuO$_3$ is a highly conducting metal in the absence of chemical substitution, and it is well lattice-matched to SrTiO$_3$ (SrTiO$_3$ = 3.91 Å, pseudocubic SrRuO$_3$ = 3.93 Å).[11] Furthermore, the interface between SrRuO$_3$ and SrTiO$_3$ is free from a polar discontinuity which can be a significant source of interface states.[12] Therefore, among the range of perovskite heterojunctions currently studied, the SrRuO$_3$/Nb:SrTiO$_3$ interface may be relatively electronically clean.

SrRuO$_3$/Nb:SrTiO$_3$ Schottky junctions were fabricated by pulsed laser deposition. A polycrystalline SrRuO$_3$ target was ablated with a KrF excimer laser at a repetition rate of 8 Hz onto Nb:SrTiO$_3$ (100) 0.01 wt % and 0.5 wt % doped substrates. The growth conditions were 700°C under an oxygen partial pressure of 0.1 Torr, using a laser fluency of 5 J/cm$^2$. High quality c-axis oriented single crystal thin films were confirmed by x-ray diffraction. The temperature dependent resistivity of a 100 nm SrRuO$_3$ film grown on an undoped substrate showed metallic behavior similar to previous reports.[11] Figure 1 shows room temperature junction *I-V* characteristics of 100 nm SrRuO$_3$ on Nb:SrTiO$_3$ for both concentrations, showing typical rectifying behavior. Ohmic contacts were formed by Ag epoxy and ultrasonic solder to the SrRuO$_3$ and Nb:SrTiO$_3$, respectively. The junction polarity is defined as a positive voltage applied to the SrRuO$_3$ electrode.

IPE measurements were taken using a 300 W halogen lamp, whose output was passed through a grating monochromator and focused into an optical fiber which guides the light onto the sample. The optical intensity at the output of the fiber was monitored using calibrated Si (290 nm to 1100 nm) and Ge (700 nm to 1800 nm) photodiodes. Before entering the fiber, the optical path passed through a chopper, and the junction photocurrent was detected by a lock-in amplifier with respect to the chopper.

Figure 2(a) shows the photon energy dependence of the square root of the photoyield $\sqrt{Y}$, defined as the photocurrent per incident photon. Two principle features can be seen in the wide scan spectra. Above 3.2 eV, the bandgap of SrTiO$_3$, $\sqrt{Y}$ sharply increases, corresponding to the photogeneration of electron-hole pairs in Nb:SrTiO$_3$. The second feature is the gradual onset of $\sqrt{Y}$ between 1.3 and 2.4 eV (Figure 2(b)). The onset of $\sqrt{Y}$ at and above the SBH is given by the Fowler formula[13] as

$$\sqrt{Y} \propto \left(h\nu - \Phi_{SB}^{IPE}\right) \qquad \left(h\nu - \Phi_{SB}^{IPE} > 3kT\right) \qquad (1)$$

where $\Phi_{SB}^{IPE}$ is the SBH as measured by IPE and $h\nu$ is the incident photon energy. The linear response obtained over a wide spectral range strongly suggests that current generation is due to electron excitations from the metal to the semiconductor, and not from excitation of electrons from localized in-gap states, which would significantly deviate from linearity. In addition, illumination of the samples from the metal side supports the dominance of the metal-semiconductor photocurrent process. The alternative geometry with junctions illuminated from the semiconductor side requires light penetration through a large volume of the substrate, increasing the probability of in-gap photocurrent generation in the semiconductor.

By extrapolating the linear portion of $\sqrt{Y}$ with respect to the incident photon energy, the barrier height is obtained. This extrapolation gives the SBH $\Phi_{SB}^{IPE}$ of 1.47±0.01 eV and 1.31±0.01 eV for Nb = 0.01 wt % and 0.5 wt %, respectively. The deviation from the straight line at low photon energies is attributed to electrons thermally excited above the Fermi level, which is neglected in the derivation of Equation (1). The experimentally determined values for $\Phi_{SB}^{IPE}$ can be compared with the SBH estimated from the Schottky-Mott model ($\Phi_{SB}^{SM}$).[14] The work function of SrRuO$_3$ is 5.2 eV,[15] and the electron affinity of SrTiO$_3$ is 3.9 eV,[16] giving $\Phi_{SB}^{SM}$ = 1.3 eV, which is in relatively good agreement with the experimentally determined $\Phi_{SB}^{IPE}$, particularly considering the simplicity of the Schottky-Mott framework. This is evidence that the formation of the SBH in SrRuO$_3$/Nb:SrTiO$_3$ is likely dominated by the electron affinity rule.

We compare these values with an independent measure of the SBH obtained from *C-V*

measurements of the junctions. In Figure 3, $C^{-2}$ versus $V$ is shown for the two junctions. For Nb = 0.01 wt %, a linear relation is observed, and the built-in potential $V_{bi}$ deduced from the voltage intercept is 1.35±0.01 V. The uncompensated donor concentration $N_D$ evaluated from the slope of the plot is $1.6 \times 10^{17}$ cm$^{-3}$, which is smaller by more than an order of magnitude in comparison with the nominal value of $3.3 \times 10^{18}$ cm$^{-3}$. Bulk Hall effect measurements of the substrate gives $N_D$ of $1 \times 10^{18}$ cm$^{-3}$, implying uniform (given the linearity of $C^{-2}$) partial passivation of the near surface donors. In order to rule out the possibility of surface passivation of the substrate surface by exposure to the growth conditions, we have studied Au/Nb:SrTiO$_3$ Schottky junctions as a function of preannealing the substrate before *in situ* Au deposition. For samples preannealed in oxygen partial pressures ranging from $1 \times 10^{-1}$ to $1 \times 10^{-7}$ torr (30min, 850°C), or with no preannealing, we find no systematic variation in $N_D$ derived from C-V measurements. We suspect the deviation originates from fluctuations of $N_D$ in the substrates, as well as the polishing process.

In the case of Nb = 0.5 wt %, nonlinear behavior is observed in $C^{-2}$ versus $V$, which can be attributed to the electric field dependence of the permittivity $\varepsilon(E)$ in SrTiO$_3$.[10] Taking $\varepsilon(E) = b/\sqrt{a+E^2}$ where $a$ and $b$ are temperature dependent constants, as established by Yamamoto *et al.*,[17] the junction capacitance can be rederived as

$$\frac{1}{C^2} = \frac{2\sqrt{a}}{b\varepsilon_0 q N_D}(V_{bi} - V) + \frac{1}{b^2 \varepsilon_0^2}(V_{bi} - V)^2, \qquad (2)$$

where $\varepsilon_0$ is the vacuum permittivity and $q$ is the electronic charge. Fitting Equation (2) to Nb = 0.5 wt % C-V data, and taking a and b parameters from Ref. 17, $V_{bi}$ of 1.40±0.1 V and $N_D$ of $2.7\times10^{20}$ cm$^{-3}$ are obtained, in reasonable agreement with the nominal value $1.7\times10^{20}$ cm$^{-3}$ and the bulk Hall effect value of $2.9\times10^{20}$ cm$^{-3}$.

In the case of a non-degenerate semiconductor Schottky junction, $V_{bi}$ should be corrected by the energy difference between the conduction band minimum and the Fermi level in the semiconductor (ξ) in order to obtain the SBH.[14] However, in the case of a degenerate semiconductor, the correction term is modified for the electrons residing in the conduction band where the band is bent, where the details of the correction depend on the density of states.[18] For SrTiO$_3$, even the undoped substrates exhibit a chemical potential pinned to the bottom of the conduction band up to room temperature, and both Nb concentrations examined here are effectively in the degenerate regime, as seen by the temperature independent Hall coefficient, for example. If we apply a parabolic band approximation for the conduction band in Nb:SrTiO$_3$, the correction term is reduced to 0.4ξ compared with the non-degenerate case.[19] Since the calculated ξ is 0.8 (117) meV for 0.01 wt % (0.5 wt %), assuming a single parabolic band with an effective mass of $m^* = 1.3m_0$ (Ref. 20), and using $N_D$ deduced from the C-V slope, the SBHs ($\Phi_{SB}^{CV}$) are calculated as 1.35±0.01 eV (1.35±0.1 eV) for Nb = 0.01 wt % (0.5 wt %) concentrations.

The SBH values deduced from the two techniques, IPE and C-V, are in reasonable agreement with one another; $\Phi_{SB}^{IPE}$ is 9 % larger than $\Phi_{SB}^{CV}$ for 0.01 wt % Nb, and 3 % smaller for 0.5 wt % Nb. Indeed, the independent measure of the SBH by IPE gives considerable support for the quadratic model for $C^{-2}$ used to incorporate dielectric nonlinearities in the 0.5 wt % Nb:SrTiO$_3$ junction – a linear extrapolation would significantly underestimate the SBH. The remaining discrepancies may arise from interface states,[6] although the detailed interface electronic structure is not currently known.

By contrast to the relative agreement of the SBH in the above analysis, extraction of the SBH from *I-V* characteristics requires more caution. Assuming transport by thermoionic emission, $\Phi_{SB}^{IV}$ is evaluated as 0.72 (0.70) V, and the ideality factor *n* as 1.53 (1.81) for Nb = 0.01 wt % (0.5 wt %). Here a Richardson constant of 156 A K$^{-2}$ was used.[20] The larger than unity *n* and $\Phi_{SB}^{IV}$ smaller than $\Phi_{SB}^{IPE}$ and $\Phi_{SB}^{CV}$ have been observed frequently in conventional semiconductor devices.[6,7,14] Origins may include a tunneling contribution to the current, spatial barrier inhomogeneities, and interface states between the metal and semiconductor. From the perspective of electrical contacts to dielectric materials, Poole-Frenkel detrapping can occur at the interfaces between transition metal oxides. However, the contribution of trap states and/or an insulating interfacial layer is regarded as small in our system, as the *C-V* results are well described by models based on direct metal-semiconductor contacts, as discussed above. The limited range of exponential behavior in these junctions near room temperature makes it difficult to deduce the primary contributions to *n* > 1, although similar junctions formed with manganite electrodes appear to be well described by thermoionic field emission.[21]

The difficulties in directly applying semiconductor transport analysis to these SrRuO$_3$/Nb:SrTiO$_3$ heterojunctions highlights the value of IPE as an independent probe for direct determination of the SBH, in the unbiased junction. This technique should be highly useful for characterizing oxide interfaces, given the emerging interest in creating and probing unusual interface charge states in perovskites.

We thank S. Nishiki for discussions and assistance. Y.H. acknowledges support from the Grant-in-Aid 21st Century COE Program at the University of Tokyo.

Figure 1

Current-voltage characteristics for SrRuO$_3$/Nb:SrTiO$_3$ junctions at room temperature. Circles (triangles) denote Nb = 0.01 wt % (0.5 wt %).

Figure 2

(a) Internal photoemission spectra of SrRuO$_3$/Nb:SrTiO$_3$ junctions at room temperature. The square root of the photoyield is plotted against the photon energy. (b) Magnified spectra of the SrRuO$_3$/Nb:SrTiO$_3$ junctions, where the solid lines denote the linear extrapolations used to extract the barrier height.

Figure 3

Capacitance-voltage characteristics of SrRuO$_3$/Nb:SrTiO$_3$ junctions plotted as $C^{-2}$ versus voltage. For Nb = 0.01 wt %, the solid line denotes the linear extrapolation used to extract $V_{bi}$. For Nb = 0.5 wt %, the solid line denotes the quadratic fit used to extract $V_{bi}$ (see text for details).

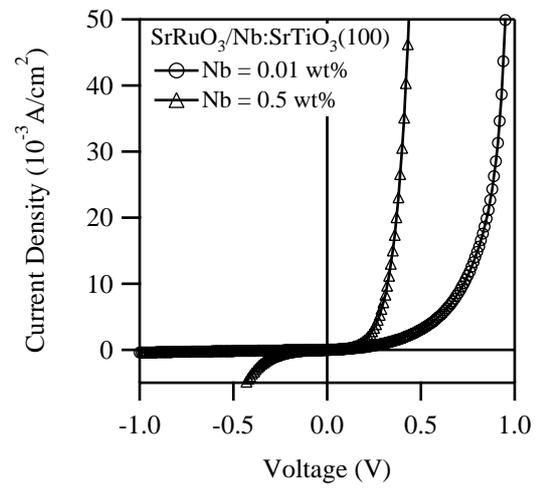

Fig. 1　Y. Hikita *et al.*

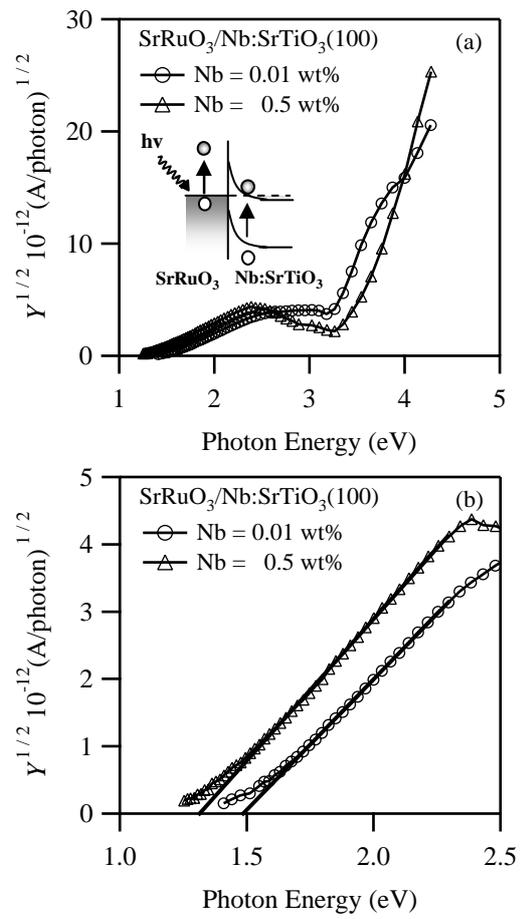

Fig. 2 Y. Hikita *et al.*

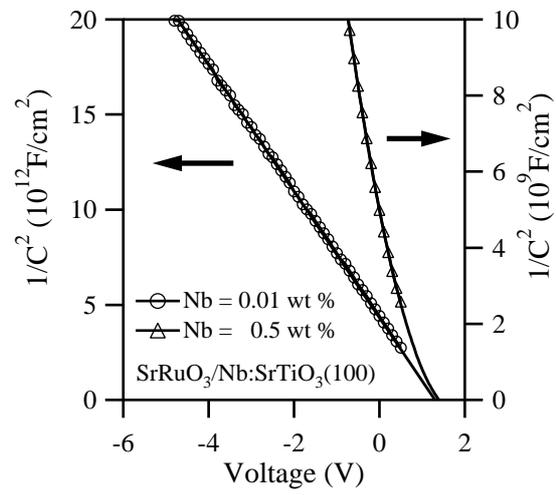

Fig. 3  Y. Hikita *et al.*